\documentclass{ws-ijmpd}
\usepackage[super,compress]{cite}
\begin{document}

\markboth{Chikun Ding, Changqing Liu and Qian Guo } {Spacetime noncommutative
effect
 on black hole as particle accelerators}

%
\catchline{}{}{}{}{}
%

\title{Spacetime noncommutative effect
 on black hole as particle accelerators  }

\author{Chikun Ding}

\address{Department of Physics and Information
Engineering, Hunan University of Humanities, Science and Technology, Loudi,
Hunan 417000, P. R. China \\
dingchikun@163.com}

\author{Changqing Liu and Qian Quo}

\address{Department of Physics and Information
Engineering, Hunan University of Humanities, Science and Technology, Loudi,
Hunan 417000, P. R. China\\
lcqliu2562@163.com}

\maketitle

\begin{history}
\received{Day Month Year}
\revised{Day Month Year}
\end{history}

\begin{abstract}
We study the spacetime noncommutative effect on black hole as particle
accelerators and, find that particle falling from infinity with zero velocity
cannot collide with unbound energy when the noncommutative Kerr black hole is
exactly extremal.  Our results also show that the bigger of the spinning
black hole's mass is, the higher of center of mass energy that the particles
obtain. For small and medium noncommutative Schwarzschild black hole, the
collision energy depends on the black holes' mass.
\end{abstract}


\ccode{PACS numbers: 04.50.Kd, 04.70.Dy, 95.30.Sf, 97.60.Lf}


\section{Introduction}
Ba\~{n}ados {\it et al} \cite{banados} and some other authors
 \cite{jacobson,wei,grib,zaslavskii,harada1} recently showed that particles falling freely
  from rest outside a Kerr black hole can collide with arbitrarily high
   center of mass  energy in the limiting case of maximal  black hole spin.
   They proposed that this might lead to signals from
  ultra high energy collisions, for example of dark matter particles.
It seems that this arbitrarily high
   center of mass  energy was generally accepted, however, it was criticized
   recently \cite{berti,jacobson,bejger,harada}. The criticism related to
   very much issues, such as (1) an infinite time being taken to
access the infinite collision energy in the extremal black hole case; (2) an
infinite narrow strip between a horizon and a potential barrier where a
particle can acquire the critical angular momentum due to multiple scattering
in the nonextremal black hole case; (3) some astrophysical limitations  such
as gravitational radiation, backreaction, etc..
  Jacobson {\it et al}
\cite{jacobson}  pointed out that ultra-energetic collisions cannot occur in
nature due to some practical limitations. Bejger {\it et al} and Harada {\it
et al} \cite{bejger,harada} pointed out that while the particle energy
diverges, the position of the collision makes it impossible to escape to
infinity and, one shouldn't expect collisions around a black hole to act as
spectacular cosmic accelerators.

 In this paper we aim to show
that when the quantum effect of gravity is considered, the presence of
infinite collision energy cannot occur. The Planck energy scale is the realm
where the general relativity will encounter with the quantum mechanics
\cite{camelia,camelia1,camelia2}. So to the energetic collision particles,
one should consider their quantum effect.

In the absence of a full quantum gravity theory, one usually uses effective
theories to describe the quantum gravitational behavior such as Quantum Field
Theory in Curved Spacetime. Noncommutative geometry is recently used as an
effective tool for modelling the extreme energy quantum gravitational effects
of the final phase of black hole evaporation, which are plagued by
singularities at a semiclassical level. Quantum mechanics teach us that the
emergence of a minimal length is a natural requirement when quantum features
of phase space are considered. It also holds true to spacetime
\cite{camelia1,garattini}. The singularities in general relativity and
ultraviolet divergences in quantum field theory can be avoided by the
presence of spacetime minimal length. These singularities and divergences are
nothing but pseudo effects due to the inadequacy of the formalism at small
scales/extreme energies, rather than actual physical phenomena.

At very short distances the classical concept of spacetime should give way to
a somewhat fuzzy picture \cite{camelia1}. The fundamental notion of the
noncommutative geometry is that the picture of spacetime as a manifold of
points breaks down at distance scales of the order of the Planck length:
Spacetime events cannot be localized with an accuracy given by Planck length
\cite{akofor}  as well as particles do in the quantum phase space. So that
the points on the classical commutative manifold should then be replaced by
states on a noncommutative algebra and the point-like object is replaced by a
smeared object \cite{smail} to cure the singularity problems at the terminal
stage of black hole evaporation \cite{nico}.

The approach to noncommutative quantum field theory follows two paths: one is
based on the Weyl-Wigner- Moyal *-product and the other on coordinate
coherent state formalism \cite{smail}. In a recent paper \cite{morita},
following the coherent state approach, it has been shown that Lorentz
invariance and unitary, which are controversial questions raised in the
*-product approach, can be achieved by
assuming $\vartheta^{\mu\nu}=\vartheta \; \text{diag}(\epsilon_1,\ldots,
\epsilon_{D/2}),$ where $\vartheta$ is a constant which has the dimension of
$length^2$, $D$ is the dimension of spacetime \cite{smail2} and, there isn't
any UV/IR mixing. Inspire by these results, various black hole solutions of
noncommutative spacetime have been found \cite{Nicolini:2008aj,smailagic}. In
this letter we use the noncommutative Kerr solution to study the problem of
the particles' center of mass energy when they collide at the horizon.

\section{ The noncommutative Kerr black hole}

In history, the study on rotating black hole solution had been met with
technical difficulties in solving Einstein equations, and completely ignored
the appropriate matter source. The obtainment of Kerr solution is based on
the so-called ``vacuum solution" method consisting in assuming an additional
symmetry for the metric and solving field equations with no matter source.
Integration constants are then determined comparing the weak field limit of
the solution with known Newtonian-like forms. This approach is physically
unsatisfactory especially in General Relativity, where basic postulate is
that geometry is determined by the mass-energy distribution. Using the
noncommutative geometry method and following the basic Einstein's idea that
spacetime is curved due to the presence of matter, Smailagic {\it et al}
derives
 the line element of the
noncommutative Kerr black hole \cite{smailagic}
\begin{eqnarray}
&&ds^2 = -\Big(1-\frac{2Mr}{\rho^2}\Big)\,dt^2
-\frac{4Mra\sin^2\theta}{\rho^2}dtd\phi+
\frac{\rho^2dr^2}{\Delta}\nonumber\\
&&\qquad + \rho^2 d\theta^2+
\Big(r^2+a^2+\frac{2Mra^2\sin^2\theta}{\rho^2}\Big)\sin^2\theta d\phi^2\,,
\label{metric}
\end{eqnarray}
and
\begin{eqnarray}
\label{sol1} M&=&\frac{2M_0}{\sqrt{\pi}}\, \gamma(\frac{3}{2} \ ,
\frac{r^2}{4\vartheta}\,),\;\;\gamma(\frac{3}{2} \ , x\,)\equiv \int^x_0
t^{1/2}e^{-t}dt,\nonumber\\ \rho^2&=&r^2+a^2\cos^2\theta,\;\;
\Delta=r^2-2Mr+a^2,
\end{eqnarray}
where $\vartheta$ is a spacetime noncommutative parameter\footnotemark
\footnotetext{The notation $\vartheta$ used here is a constant as well as
Plank constant $\hbar$, but we still call it a spacetime noncommutative
parameter since it up to now is undetermined.}, $a$ is the spinning black
hole's angular momentum. The commutative Kerr metric is obtained from
(\ref{metric}) in the limit $r/\sqrt{\vartheta}\to\infty $. Equation
(\ref{metric}) leads to the mass distribution $M\left(\, r\,\right)= 2M_0
\,\gamma\left(3/2\ , r^2/4\vartheta\, \right)/\sqrt\pi $, where $M_0$ is the
total mass of the source. In the classical General
  Relativity, black hole's mass is dealt by point-like mass, and then it can be
  a constant. But in the noncommutative gravity, the mass cannot be treated as
  a constant, it is the mass distribution $M(r)$.

Depending on the values of $a$, $\sqrt{\vartheta}$ and $M_0$, the metric
displays different causal structure: existence of two horizons (non-extremal
black hole), one horizon (extremal black hole) or no horizons (massive
spinning ¡°droplet¡± ). Due to $\Delta(r_+)=0$ cannot be solved analytically,
we list some values of the maximum angular momentum $a_{\rm max}$, the single
horizon $r_+$ and mass distribution $M_+$ on the horizon in Table
\ref{tabel0} by letting $M_0=1$\footnotemark \footnotetext{The units we used
here and hereafter is the total mass of the black hole $M_0$, i.e.,
$\frac{r}{M_0}\rightarrow r, \;\frac{a}{M_0}\rightarrow a,\;
\frac{\sqrt{\vartheta}}{M_0}\rightarrow\sqrt{\vartheta}.$ }.

\begin{table}[!h] \tbl{Numerical values for the
radius of the single event horizon and the mass distribution on the horizon
in the extremal spinning noncommutative black hole spacetime with different
$\sqrt{\vartheta}$ and $a_{\rm max}$ ($M_0=1$).}
{\begin{tabular}{@{}ccccccccc@{}} \toprule $\sqrt{\vartheta}$
&0.525177&0.52517&0.525&0.52 & 0.48 &0.44& 0.40&0.36 \\
 \colrule
$a_{\rm max}$& 0&0.00589&0.029159&0.15742&0.45983&0.62170&0.73892&0.82841\\
 $r_{+}$&1.58826&1.58749&1.58727& 1.58460&1.54842&1.49613&1.43401&1.36328\\
 $M_+$&0.79413&0.79376&0.79390& 0.80012&0.84287&0.87724&0.90738&0.93336\\
 $\sqrt{\vartheta}$&0.32 &0.28&0.24&0.20& 0.16 &0.12&
0.08&0.04 \\
 $a_{\rm max}$&0.89656&0.94621&0.97876&0.99539&0.99979&0.9999998&
 $1-10^{-14}$&1.00000\\
 $r_{+}$&1.28295&1.20207&1.11883&1.04683&1.00638& 1.00035& $1.000000141
 $&1.00000\\
 $M_{+}$&0.95475&0.97344&0.987528&0.99665&0.99981& $1-10^{-7}$& $1.00000
 $&1.00000\\ \botrule
\end{tabular} \label{tabel0}}
\end{table}

Table \ref{tabel0} shows that the maximum angular momentum $a_{\rm max}$
decreases with the increase of the spacetime noncommutative parameter
$\sqrt{\vartheta}$. It indicates the restriction of the spacetime
non-commutativity on the angular momentum of black hole which implies that i)
if $\sqrt{\vartheta}$ is strong, its single horizon is close to that of the
noncommutative Schwarzschild black hole; ii) if $\sqrt{\vartheta}$ is weak,
its single horizon is close to that of the commutative Kerr hole. In other
words, the point-like structure of spacetime lets $a\leq M_0$, while the
minimal length of spacetime leads to $a<M_0$.

 When
$M_0>1.90412\sqrt{\vartheta}$ and $0\leq a<a_{\rm max}$, the two horizons
(non-extremal black hole) are given by
\begin{eqnarray}
r_\pm^2=\frac{4r_\pm}{\sqrt{\pi}}\,\gamma\left(3/2\ , r^2_\pm/4\vartheta\,
\right)-a^2.
\end{eqnarray}
which is different from the commutative Kerr black hole. The line element
(\ref{metric}) describes the geometry of a noncommutative black hole and
should give us useful insights about possible spacetime noncommutative
effects on particle accelerators.

\section{Near horizon collision in extremal noncommutative Kerr black hole
spacetime }

The solution to the geodesic equation of the noncommutative Kerr black hole
is given by
\begin{equation}
  \left.
   \begin{aligned}[c]
 &\frac{dt}{d\tau}=-\frac{2MraL+a^2E\Delta-E(r^{2}+a^{2})^2}{r^{2}\Delta},
 \label{4S}\\
 &\frac{dr}{d\tau}=\pm\frac{\sqrt{2Mr(L-aE)^2+2Mr^3E^2-r^2L^2
 +\Delta r^2(E^2-m^2)}}{r^{2}},\\
 &\frac{d\phi}{d\tau}=-\frac{(a^2-\Delta)L-2MraE}{r^{2}\Delta},
    \end{aligned}
  \right.
\end{equation}
where $E,\;L,\;m$ are the particle's energy, angular momentum and rest mass.
We assume throughout the paper that the motion of particles occur in the
equatorial plane.

\begin{figure}[ht]
\begin{center}
 \includegraphics[width=6.0cm]{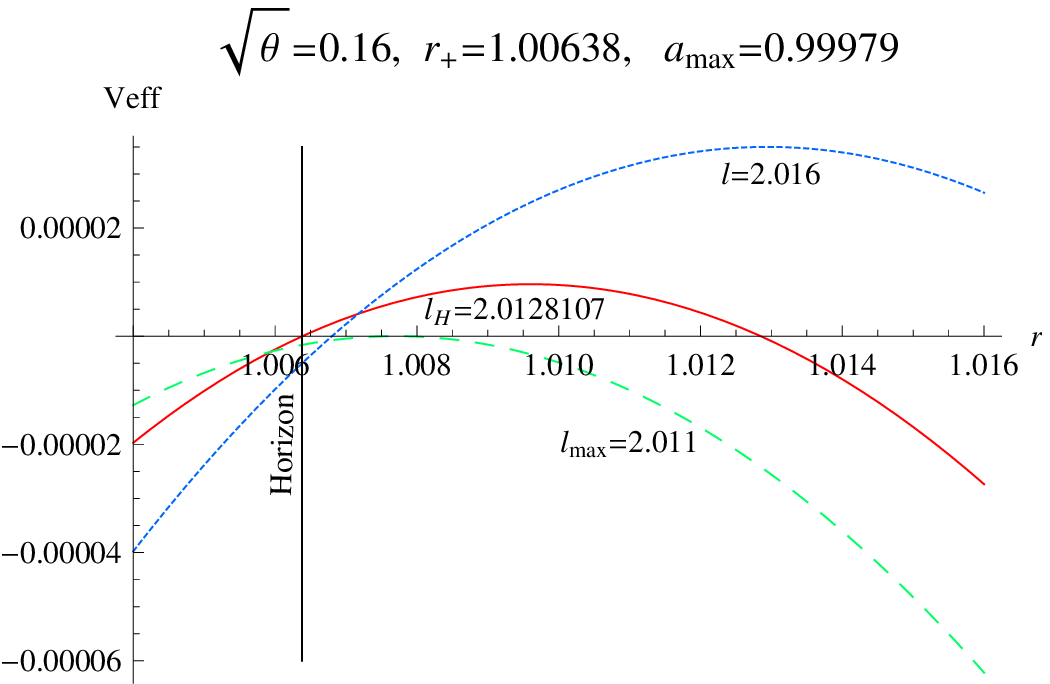}
 \includegraphics[width=6.0cm]{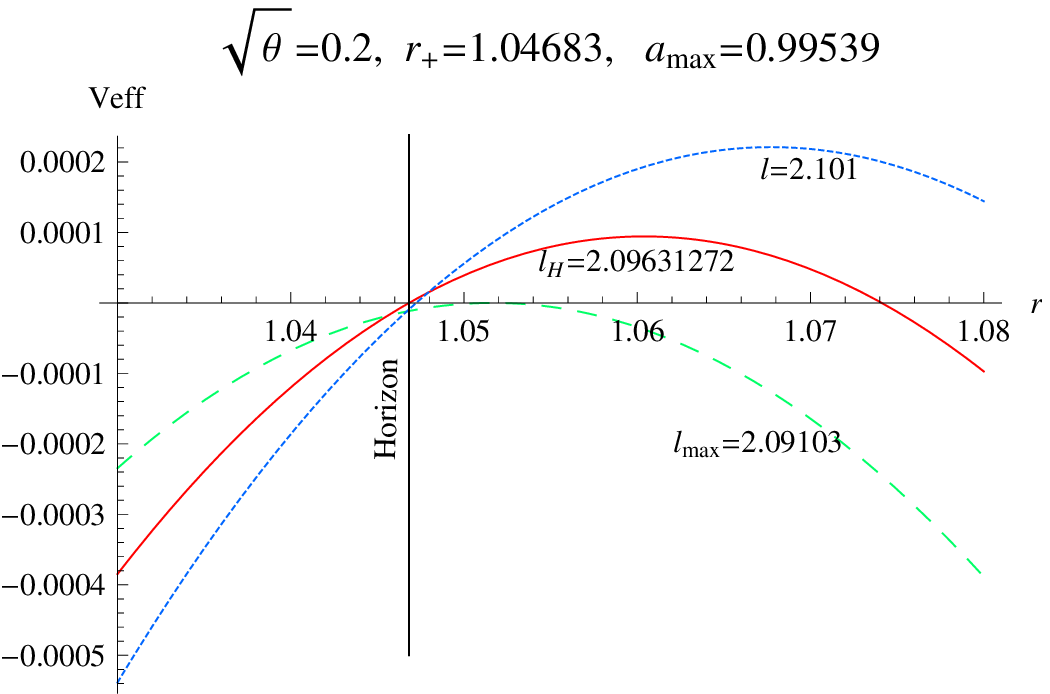}\\
 \includegraphics[width=6.0cm]{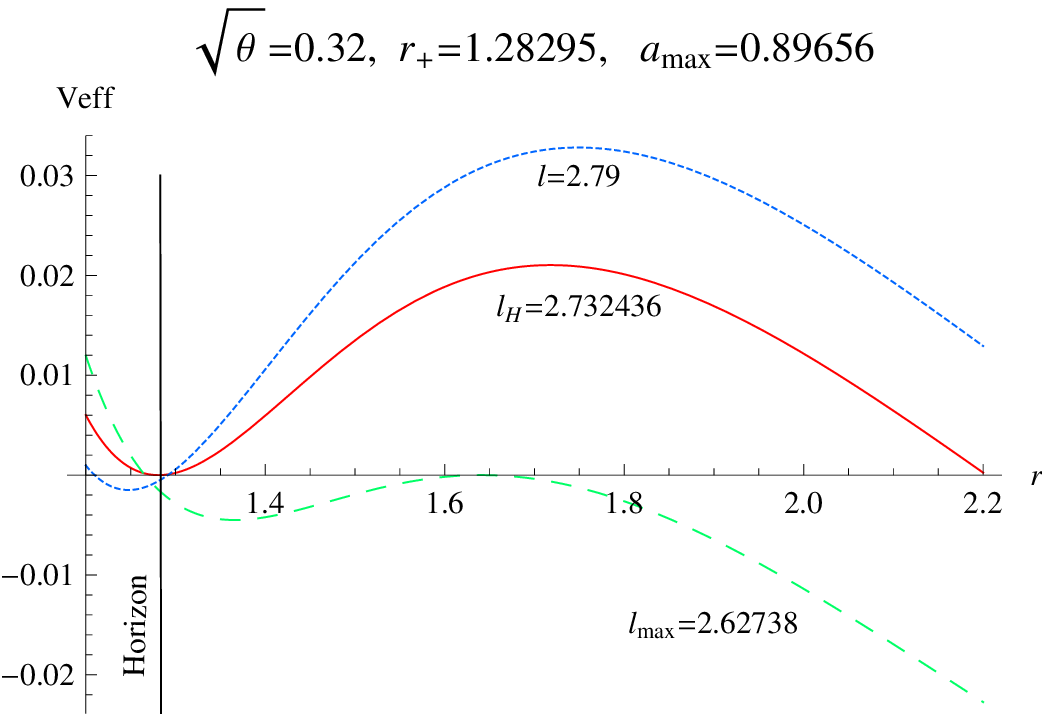}
 \includegraphics[width=6.0cm]{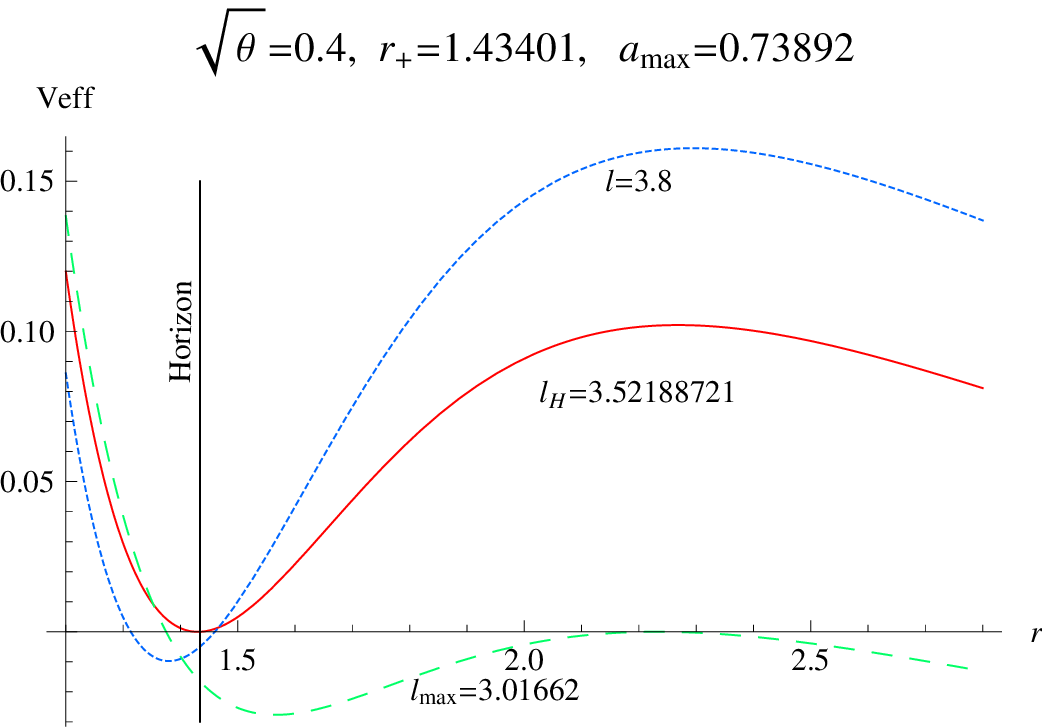}
\caption{The effective potentials of particles with critical, super-critical
and maximum angular momentum in the extremal spinning noncommutative black
hole spacetime with different $\sqrt{\vartheta}$ and $a_{\rm max}$ with
$M_0=1, m=1, E=1$.} \label{f2}
\end{center}
 \end{figure}

 Firstly, we should find the range of angular momentum
 of particles which can reach to the horizon
under the condition $a=a_{\rm max}$. The maximum/minimum angular momentum
 of particles can be found using the effective
potential for the radial motion in the equatorial plane. The proper time
derivative of the (Boyer-Lindquist) radial coordinate of orbital motion
satisfies $ \dot{r}^2/2 + V_{\rm eff}(r,L,\sqrt{\vartheta})=0, $
 where the effective potential
is given in terms of the angular momentum $L$ by
 \begin{equation} V_{\rm eff}=
-\frac{Mm^2}{r}+\frac{L^2-a^2(E^2-m^2)}{2r^2} -\frac{M(L-aE)^2}{r^3}
-\frac{E^2-m^2}{2}.
\end{equation} The maximum/minimum angular momentum we are looking for is
defined by
 $ V_{\rm eff}=
dV_{\rm eff}/dr= 0.
 $ The numerical values of maximum/minimum angular momentum are listed in
 Table \ref{tabel1}, and some effective potentials of particles with the
  critical, super-critical and maximum angular momentum are showed in Fig.
  \ref{f2}.

Secondly, we should find the critical angular momentum of particles whose
center of mass energy $E_{\rm cm}$ were assumed to be arbitrary high when
$a=a_{\rm max}$. On the background metric (\ref{metric}), the CM energy of
two particles 1 and 2 is\cite{harada1}
\begin{eqnarray}
 \frac{E^2_{\text{cm}}}{2}&=&m^2+E_1E_2+\frac{F(r)-G(r)}{D(r)},
 G(r)=2\sqrt{-V_{{\rm eff} 1}}\sqrt{-V_{{\rm eff} 2}},
 D(r)=\frac{a^2}{r^2}-\frac{2M}{r}+1,
  \nonumber\\F(r)&=&2\big[\frac{a^2}{r^2}(1+\frac{M}{r})+(1-\frac{M}{r})\big]
  E_1E_2-\frac{2Ma}{r^3}(E_1L_2+E_2L_1)-(1-\frac{2M}{r})\frac{L_1L_2}{r^2},
  \label{energy0}
\end{eqnarray}
where two particles' mass $m_1=m_2=m$. It is believed that if the collision
occurs near the horizon, the CM energy can be unboundedly high. So the
behavior of formula (\ref{energy0}) near horizon should be considered. Here
we would like to use Zaslavskii's formula \cite{zaslavskii} for seeking
$E_{\rm cm}$ in a model-independent form
\begin{eqnarray}
 \Big(\frac{E^2_{\text{cm}}}{2m^2}\Big)_H&=&1+\frac{b_{1H}(L_{2H}-L_2)}{2(L_{1H}-L_1)}
 +\frac{b_{2H}(L_{1H}-L_1)}{2(L_{2H}-L_2)}-\frac{L_{1}L_{2}}{(g_{\phi\phi})_H},\;\;
  \nonumber\\L_{iH}&=&\frac{E_i}{\omega_H},\;\;b_{iH}=1+\frac{L_{iH}^2}{(g_{\phi\phi})_H},
  \label{energy}
\end{eqnarray}
where $ \omega_H=(-g_{t\phi})_H/(g_{\phi\phi})_H$, $L=lmM_0$. After taking
$E_1=E_2=E$ for simplicity, we obtain the center of mass energy for the
collision:
\begin{eqnarray}\label{ecm}
 \Big(\frac{E_{\text{cm}}}{2m}\Big)_H=\sqrt{1+\frac{b_H(l_1-l_2)^2}
 {4(l_H-l_1)(l_H-l_2)}}\;.
\end{eqnarray}
 Then the critical angular momentum $l_H$ of particles whose center of mass
energy $E_{\rm cm}$ were assumed to be arbitrary high when $a=a_{\rm max}$
can be found via
\begin{eqnarray}
l_H=\frac{E}{\omega_H}=\frac{r_+(r_+^2+a^2)+2M_+a^2}{2M_+a}.
\end{eqnarray}
 The numerical values of critical
angular momentum are listed in Table \ref{tabel1}.

\begin{table}[!h] \tbl{Numerical values for the
maximum, minimum and critical angular momentum in the extremal spinning
noncommutative black hole spacetime with different $\sqrt{\vartheta}$ and
$a_{\rm max}$ with $M_0=1, m=1, E=1$. }
 {\begin{tabular}{@{}ccccccccc@{}} \toprule $\sqrt{\vartheta}$ &0.525177&0.52517&0.525&0.52 & 0.48 &0.44&
0.40&0.36   \\ \colrule
$l_{\rm max}$& 4.0&3.99409&3.97061&3.83578&3.46907&3.22759&3.01662&2.81895\\
 $l_{\rm min}$&$-$4.0& $-$4.00587&$-$4.02894&$-$4.15167&$-$4.41647&$-$4.54692&$-$4.63736&$-$4.70437\\
 $l_{\rm H}$&$\infty$&427.8693&86.4322&16.1080528&5.6739587&4.222171&3.52188779
 &3.07190361\\
 $\sqrt{\vartheta}$ &0.32&0.28&0.24&0.20& 0.16 &0.12&
0.08&0.04 \\
 $l_{\rm max}$&2.62738&2.4382&2.24999&2.09103&2.011& 2.0005&2.0000002&2.00000\\
 $l_{\rm min}$&$-$4.75431&$-$4.79013&$-$4.81337&$-$4.82517&$-$4.82828&$-$4.82843&$-$4.82843&$-$4.82843\\
 $l_{\rm H}$&2.732436&2.47333205&2.2576964&2.09631272&2.0128107&2.0007053&2.00000028&2.00000\\
 \botrule
\end{tabular} }\label{tabel1}
\end{table}

From Table \ref{tabel1}, one can see that all the critical angular momentum
lies beyond the range $(l_{\rm min},\; l_{\rm max})$, which shows that the
unlimited center of mass energy cannot be approached. In addition, it is
interesting that $l_{\rm min}=-l_{\rm max}=-4.0,\;l_H=\infty$ when
$\sqrt{\vartheta}$ is maximum, which is the same as that of the commutative
Schwarzschild black hole; $l_{\rm min}=-4.82843,\;l_{\rm max}=l_H=2.0$ when
$\sqrt{\vartheta}\rightarrow0$, which is the same as that of the commutative
Kerr black hole. Fig. \ref{f2} shows that the particle effective potentials
with critical angular momentum is positive near the horizon, so they cannot
approach to the horizon.

 With these data, we obtain $E_{\rm cm}$ for noncommutative
Kerr black holes with $l_1=l_{\rm min},\;l_2=l_{\rm max},\;a=a_{\rm max}$  by
using Eq. (\ref{ecm}) with
\begin{eqnarray}
b_H=1+\frac{l_{H}^2}{(g_{\phi\phi})_H}=1+\frac{r_+\Big[r_+(r_+^2+a^2)+2M_+a^2\Big]}
{(2M_+a)^2},
\end{eqnarray}
and list them in Table \ref{tabel3}.

\begin{table}[!h]
\tbl{Numerical values for the center of mass energy of particles colliding at
the horizon in the extremal spinning noncommutative black hole spacetime with
different $\sqrt{\vartheta}$ and $a_{\rm max}$ with $M_0=1, E=m=1$.}
{\begin{tabular}{@{}ccccccccc@{}} \toprule $\sqrt{\vartheta}$
&0.525177&0.52517&0.525&0.52 & 0.48 &0.44& 0.40&0.36
 \\ \colrule
$E_{\rm cm}$& 5.41948&5.42186&5.42481&5.50061&6.20380&7.14606&8.46939&10.5215\\
 $\sqrt{\vartheta}$&0.32 &0.28&0.24&0.20& 0.16 &0.12&
0.08&0.04 \\
 $E_{\rm cm}$&14.6674&23.2222&46.0119&52.5201&87.1905&258.789&12962.9&$\infty$\\
 \botrule
\end{tabular} \label{tabel3}}
\end{table}

From Table \ref{tabel3}, one can see that, for the noncommutative Kerr black
hole case, the center of mass energy of particles is bounded by using a more
generic formula in a model-independent form.

Here we can see the spacetime noncommutative effect on black hole as particle
accelerators. The spacetime noncommutative effects avoid  the presence of
infinite collision energy via preventing the black hole's angular momentum to
reach to the black hole's mass, $a=M_0$. In other words, the point-like
structure of spacetime lets $E_{\rm cm}\leq\infty$, while the presence of
spacetime minimal length leads to $E_{\rm cm}<\infty$

From Table \ref{tabel3}, one can also see that, the bounded $E_{\rm cm}$
increases with the spacetime noncommutative parameter decreasing and, if
$\sqrt{\vartheta}\rightarrow0$, it coincides with that of the commutative
case.

If we choose the spacetime noncommutative constant $\sqrt{\vartheta}=1$
units\footnotemark\footnotetext{The units we used here
 is the spacetime noncommutative constant $\sqrt{\vartheta}$,
  i.e. $\frac{r}{\sqrt{\vartheta}}\rightarrow
r,\;\frac{a}{\sqrt{\vartheta}}\rightarrow a,\;
\frac{M_0}{\sqrt{\vartheta}}\rightarrow M_0.$}, then decrease of
$\sqrt{\vartheta}$ in Table \ref{tabel3} is corresponding to increase of
black hole mass $M_0$. It can be easily seen from Table \ref{tab5}.
$\sqrt{\vartheta}\rightarrow0$ is corresponding to $M_0\rightarrow\infty$.
Therefore $E_{cm}$ increases with the black hole's mass increasing, and it
cannot be approached to arbitrary high unless the black hole mass is
infinite. This can be easily seen from Table \ref{tab5} which is related to
Table \ref{tabel0} by take the mass parameter $M_0$ in the place of
$\sqrt{\vartheta}$.

\begin{table}[!h]
\tbl{Numerical values for the radius of the single event horizon in the
extremal spinning noncommutative black hole spacetime with different $M_0$
and $a_{\rm max}$ with $\sqrt{\vartheta}=1$. }
{\begin{tabular}{@{}ccccccccc@{}} \toprule $M_0$ &1.90412&2.3&2.7&3.1 &3.5
&3.9& 4.3&4.7
\\ \colrule
$a_{\rm max}$& 0&1.46963&2.18007&2.76750&3.29068&3.77295&4.22698&4.66100\\
 $r_{+}$&3.02343&3.42096&3.7305& 3.99299&4.24353&4.48729&4.73766&5.00025\\
 $M_0$&5.1 &5.5&5.9&6.3& 6.7 &7.1&
7.5&7.9\\
 $a_{\rm max}$&5.08100&5.49169&5.58677&6.29888&6.69965&7.09990&7.49997&7.89999\\
 $r_{+}$&5.28989&6.60919&5.95209&6.32217&6.71019&7.10614&7.50523&7.90399\\
 \botrule
\end{tabular} \label{tab5}}
\end{table}

 In Table \ref{tabel3}, when $\sqrt{\vartheta}=0.525177,\;a=0,\; l_H=\infty$,
 then Eq. (\ref{ecm}) is invalid any more. We use another formula to seek
$E_{\rm cm}$ for noncommutative Schwarzschild black holes with $l_1=l_{\rm
min},\;l_2=l_{\rm max},\;a=0$ \cite{banados}
\begin{eqnarray}
 &&\Big(\frac{E_{\rm cm}}{m}\Big)_H^2=2\frac{H'}{(r\Delta)'}\Big| r\rightarrow r_+,\;
\qquad\qquad H=\big(2M-r\big)l_1l_2-2Ma(l_1+l_2)+2r(r^2+a^2)\nonumber\\
   &&
\qquad+2M(a^2-r^2)
-\sqrt{2M(l_1-a)^2+2Mr^2-rl_1^2}\sqrt{2M(l_2-a)^2+2Mr^2-rl_2^2},
\end{eqnarray}
where the prime $'$ denotes ${d}/{dr}$.

\begin{table}[!h]
\tbl{Numerical values for the center of mass energy of particles colliding at
the horizon in the noncommutative Schwarzschild black hole spacetime with
different $\sqrt{\vartheta}$ with $M_0=1, E=m=1$.}
 {\begin{tabular}{@{}ccccccccc@{}}
\toprule $\sqrt{\vartheta}$ &0.525177&0.52517&0.525&0.52 & 0.48 &0.44&
0.40&0.36  \\ \colrule
$E_{\rm cm}$& 5.41948&5.40702&5.34855&5.07212&4.65865&4.54187&4.49482&4.47757\\
 $\sqrt{\vartheta}$&0.32 &0.28&0.24&0.20& 0.16 &0.12&
0.08&0.04 \\
 $E_{\rm cm}$&4.47290&4.47218&4.47214&4.47214&4.47214&4.47214&4.47214&4.47214\\
\botrule
\end{tabular} \label{tabel2}}
\end{table}

From Table \ref{tabel2}, one can see that, for the noncommutative
Schwarzschild black hole case, the bounded $E_{\rm cm}$ increases with the
spacetime noncommutative parameter if $0.52517<\sqrt{\vartheta}\leq0.24$
which is different from commutative case.
 If $0<\sqrt{\vartheta}<0.24$, it coincides with that of the
commutative case, i.e. the collision energy does not depend on the mass of
black hole \cite{banados}.

\section{ISCO particle collision in extremal spinning noncommutative black
hole spacetime }

Harada {\it et al} \cite{harada1} pointed out that either a particle plunging
from the ISCO (innermost stable circular orbit) to the horizon or orbiting
the ISCO collides with another particle can obtain an arbitrarily high CM
energy without any artificial fine-tuning in an astrophysical context. In
this section, we consider these collisions in the extremal spinning
noncommutative black hole spacetime.

\begin{figure}[ht]
\begin{center}
 \includegraphics[width=6.0cm]{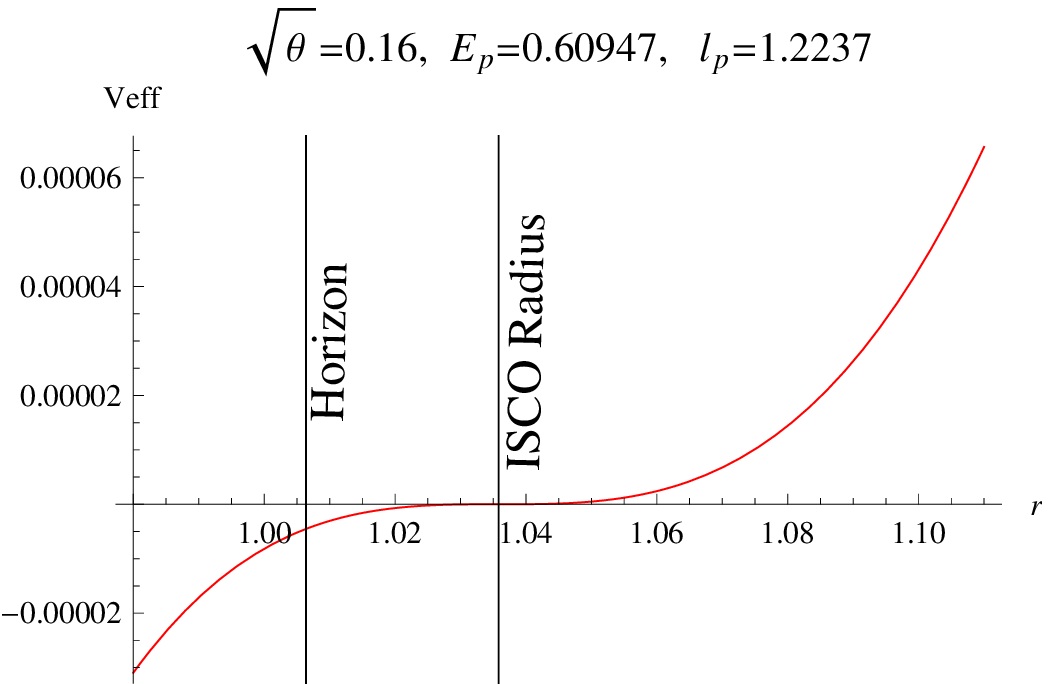}
 \includegraphics[width=6.0cm]{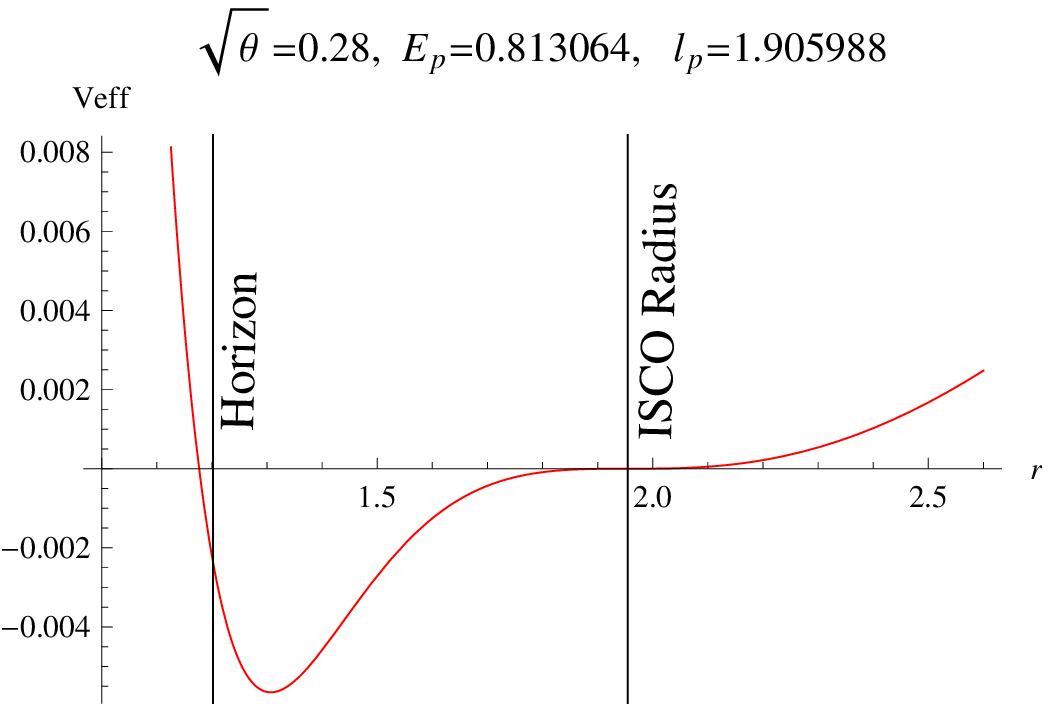}
\caption{The effective potentials of the prograde ISCO particles in the
extremal spinning noncommutative black hole spacetime with different
$\sqrt{\vartheta}$ ($M_0=1, m=1$).} \label{fisco}
\end{center}
 \end{figure}

The ISCO in the Kerr spacetime is explicitly given by Bardeen, Press and
Teukolsky\cite{bardeen}. The circular orbit on the equatorial plane is given
by $V_{\rm eff}(r)=0, V'_{\rm eff}(r)=0$, and the ISCO is determined by the
condition  $V''_{\rm eff}(r)=0$. Here we consider only the prograde ISCO, and
the numerical values for the radius of the prograde ISCO $r_p$, the particle
energy $E_p$. angular momentum $l_p$ are listed in Table \ref{tabelisco},
some effective potentials are showed in Fig. \ref{fisco}.
\begin{table}[!h] \tbl{Numerical values for the
radius of the prograde ISCO $r_p$, the particle energy $E_p$ angular momentum
$l_p$ in the extremal spinning noncommutative black hole spacetime with
different $\sqrt{\vartheta}$ and $a_{\rm max}$($M_0=1, m=1$). }
 {\begin{tabular}{@{}ccccccccc@{}} \toprule $\sqrt{\vartheta}$ &0.525177&0.52517&0.525&0.52 & 0.48 &0.44&
0.40&0.36   \\ \colrule
$r_{\rm p}$& 6.0&5.98075&5.90443&5.47538&4.38851&3.7374&3.21036&2.75103\\
 $E_{\rm p}$&0.942809&0.94262&0.94185&0.93715&0.92092&0.90644&0.89017&0.87057\\
 $l_{\rm p}$&3.4641&3.45854&3.43639&3.30874&2.95718&2.72145&2.51181
 &2.31115\\
 $\sqrt{\vartheta}$ &0.32&0.28&0.24&0.20& 0.16 &0.12&
0.08&0.04 \\
 $r_{\rm p}$&2.33639&1.9544&1.59737&1.26164&1.03582&1.00628&1.0000002&1.00000\\
 $E_{\rm p}$&0.84581&0.81306&0.76739&0.69992&0.60947&0.58099&0.5773504&0.5773503\\
 $l_{\rm p}$&2.11120&1.90599&1.68933&1.45464&1.2237&1.16201&1.1547008&1.1547007\\
 \botrule
\end{tabular} }\label{tabelisco}
\end{table}

From Table \ref{tabelisco}, we can see that when noncommutative parameter
$\sqrt{\vartheta}\rightarrow0$, the prograde ISCO radius $r_p\rightarrow
M_0$, particle energy and angular momentum $E_p\rightarrow m/\sqrt{3},
L_p\rightarrow 2mM_0/\sqrt{3}$. When noncommutative parameter
$\sqrt{\vartheta}\rightarrow0.525177$, the prograde ISCO radius
$r_p\rightarrow 6M_0$. These results coincides with those ones obtained in
commutative case.

With these data, the CM energy of an ISCO particle collision can be obtained.
As for the case that a particle plunging from the ISCO collides with another
one, we assume that $E_1=E_p, l_1=l_p, E_2=1, l_2=l_{min}$ and use the
formula (\ref{energy}) to obtain CM energy. The numerical values are listed
in Table \ref{tabelisco1}. It is easy to see that it cannot collide with
arbitrary high CM energy.
\begin{table}[!h]
\tbl{Numerical values for the center of mass energy of particle plunging from
the prograde ISCO colliding with other particle near the horizon in the
extremal spinning noncommutative Kerr black hole spacetime with different
$\sqrt{\vartheta}$ ($l_1=l_p, l_2=l_{min}, M_0=1, m=1$).}
 {\begin{tabular}{@{}ccccccccc@{}}
\toprule $\sqrt{\vartheta}$ &0.525177&0.52517&0.525&0.52 & 0.48 &0.44&
0.40&0.36  \\ \colrule
$E_{\rm cm}$&4.14213&4.15491&4.43082&4.84061&5.56238&6.93147&7.84845&8.58937\\
 $\sqrt{\vartheta}$&0.32 &0.28&0.24&0.20& 0.16 &0.12&
0.08&0.04 \\
 $E_{\rm cm}$&9.99493&12.2423&17.1973&29.1802&55.6678&155.734&7478&35178.1\\
\botrule
\end{tabular} \label{tabelisco1}}
\end{table}

As for the case that a particle orbiting on the ISCO and collides with
another one, we assume that $E_1=E_p, l_1=l_p, E_2=1, l_2=l_{min}$ and use
the formula (\ref{energy0}) with $r=r_p$ to obtain CM energy. The numerical
values are listed in Table \ref{tabelisco2}. It is easy to see that it cannot
also collide with arbitrary high CM energy.
\begin{table}[!h]
\tbl{Numerical values for the center of mass energy of particle on the
prograde ISCO colliding with other particle near the horizon in the extremal
spinning noncommutative Kerr black hole spacetime with different
$\sqrt{\vartheta}$ ($l_1=l_p, l_2=l_{min}, M_0=1, m=1$).}
 {\begin{tabular}{@{}ccccccccc@{}}
\toprule $\sqrt{\vartheta}$ &0.525177&0.52517&0.525&0.52 & 0.48 &0.44&
0.40&0.36  \\ \colrule
$E_{\rm cm}$&2.36601&2.36781&2.4048&2.42282&2.59655&2.76469&2.96751&3.2357\\
 $\sqrt{\vartheta}$&0.32 &0.28&0.24&0.20& 0.16 &0.12&
0.08&0.04 \\
 $E_{\rm cm}$&3.6945&4.33574&5.3927&8.1807&22.9037&48.9025&6606.5&30173.4\\
\botrule
\end{tabular} \label{tabelisco2}}
\end{table}

\section{Near horizon collision in nonextremal noncommutative Kerr black hole
spacetime }

For the nonextremal horizon, A particle with $E=1$ cannot penetrate from
infinity to the horizon but, nonetheless, there is a narrow region between a
horizon and a potential barrier where such motion can occur that can generate
acceleration to arbitrary large energies \cite{zaslavskii,grib}
\begin{eqnarray}
 0\leq r-r_+\leq r_{\rm max},
\end{eqnarray}
where
\begin{eqnarray}
 r_{\rm max}=\frac{\varepsilon^2}{b_H(N^2)'(r_+)},
 \;N^2=\frac{g^2_{t\phi}}
 {g_{\phi\phi}}-g_{tt}=\frac{\Delta}{r^2+a^2+2Ma^2/r},\;
 \varepsilon=1-\frac{l}{l_H}.
\end{eqnarray}
Some numerical values of $(r_{\rm max},\;E_{\rm cm})$ for noncommutative Kerr
black hole and $(r'_{\rm max},\;E'_{\rm cm})$ for commutative Kerr black hole
are listed in Table \ref{tabel4}.

\begin{table}[!h]
\tbl{Numerical values of $(r_{\rm max},\;E_{\rm cm})$ for noncommutative Kerr
black hole and $(r'_{\rm max},\;E'_{\rm cm})$ for commutative Kerr black hole
with $a=a_{\rm max}(1-0.01),\;l_1=l_H(1-0.01),l_2=l_{\rm min}$ and Eq.
(\ref{ecm}) with $M_0=1, m=1$. }
 {\begin{tabular}{@{}ccccccccc@{}} \toprule $\sqrt{\vartheta}$ &0.36&0.32&0.28&0.24 &0.20 &0.16&
0.12&0.08
\\ \colrule
$r_{\rm max}\times10^{-4}$&2.72803&3.19306&3.74738&4.47368&5.40021
 &6.03336&6.08874&6.08881\\
 $r'_{\rm
 max}\times10^{-4}$&$0.74769$&1.17096&1.85710&3.04582&4.88076&6.01662&6.08874
 &6.08881\\
$E_{\rm cm}$ &31.2842&29.5185&28.2908&27.4611&26.8900 &26.8264&
26.8180&26.8180 \\
 $E'_{\rm cm}$&32.1719&30.0384&28.5690&27.5252&27.0018&
 26.8268&26.8180&26.8180\\
 \botrule
\end{tabular} \label{tabel4}}
\end{table}

From table \ref{tabel4}, it is easy to see the region between a horizon and a
potential barrier is infinite narrow where a particle can acquire the
critical angular momentum due to multiple scattering. It is also shows that
the spacetime noncommutative effect constrains this $E_{\rm cm}$ also in
nonextremal rotating black hole spacetime.

\section{ Summary}

We have examined the mechanism that using spinning and non-spinning black
holes as particle accelerators in presence of quantum effect of gravity. Our
results show that infinite center of mass energy for the colliding particles
cannot be attained unless the mass of the black hole is infinite. This is due
to that the point-like structure of spacetime lets $ a\leq M_0, $ and $E_{\rm
cm}\leq\infty$; while the presence of spacetime minimal length leads to $a<
M_0, $ and $E_{\rm cm}<\infty$.

The present mechanisms that prevent infinite energies are (1) an infinite
time being taken to access the infinite collision energy in the extremal
black hole case; (2) an infinite narrow strip between a horizon and a
potential barrier where a particle can acquire the critical angular momentum
due to multiple scattering in the nonextremal black hole case; (3) some
astrophysical limitations  such as gravitational radiation, backreaction,
etc. So one can see that the quantum effect of gravity is an other preventing
mechanism.

 Additionally, for noncommutative rotating black hole, the collision
energy increases with the increasing of black hole's mass as the black hole
is exactly extremal. For noncommutative Schwarzschild black hole, the bound
$E_{\rm cm}$ decreases with the black hole mass if
$0.52517<\sqrt{\vartheta}\leq0.24$ which is different from commutative case
and,
 if $0<\sqrt{\vartheta}<0.24$,  the collision energy does not depend on the mass of
black hole which coincides with that of the commutative case.

Gamma-ray bursts and ultra-high-energy cosmic rays provide an important
testing ground for fundamental physics. Some cosmic rays have been observed
with extremely high energies. These rays may be generated at the black hole
horizon which acts as a particle accelerator. Our study shows that in order
to acquire higher energy, the bigger of black hole mass is required. It can
be used to explore new ideas in the structure of spacetime at short
(Planckian) distance scale.

\section*{Acknowledgments}
This work was partially supported by Hunan Provincial Natural Science
Foundation of China No. 11JJ3014, the Scientific Research Fund of Hunan
Provincial Education Department No. 11B067, the National Natural Science
Foundation of China under No. 11247013; Aid program for Science and
Technology Innovative Research Team in Higher Educational Institutions of
Hunan Province;  C. Liu's work was supported by the Hunan Provincial
Innovation Foundation for Postgraduate NO. CX2011B185.

\end{document}